\pdfoutput=1
\documentclass[11pt]{article}
\usepackage[preprint]{coling}
\usepackage{times}
\usepackage{latexsym}
\usepackage[T1]{fontenc}
\usepackage[utf8]{inputenc}
\usepackage{microtype}
\usepackage{inconsolata}
\usepackage{graphicx}
\usepackage{hhline,array,tabularx,colortbl,multirow,caption}
\usepackage{amsmath,amssymb}
\newcounter{algorithm}
\newcommand{\algline}[3]{%
  \noindent\hangindent=\dimexpr 1.50em + #2em\relax
  \makebox[1.05em][r]{#1}\hspace{0.45em}\hspace*{#2em}#3\par}

\title{Accelerating Token-and-Duration Transducer ASR Inference:\\
A Production Case Study of Parakeet TDT 0.6B}
\author{%
  \textbf{Aditya Chatterjee\textsuperscript{1}, Sanjay Singh Chauhan\textsuperscript{2}, Prabhsimran Singh\textsuperscript{1}}\\
  \textbf{Arushi Dhingra\textsuperscript{2}, Mayank Jain\textsuperscript{2}, Rahul Mittal\textsuperscript{2}}\\
  \textsuperscript{1}NVIDIA, Bangalore, India \qquad
  \textsuperscript{2}NVIDIA, Pune, India\\
  \small\texttt{\{adichatterje, schauhan, prabhsimrans, arushid, mayjain, rmittal\}@nvidia.com}}
\hypersetup{pdftitle={Accelerating Token-and-Duration Transducer ASR Inference: A Production Case Study of Parakeet TDT 0.6B},
  pdfauthor={Aditya Chatterjee, Sanjay Singh Chauhan, Prabhsimran Singh, Arushi Dhingra, Mayank Jain, Rahul Mittal},pdfkeywords={ML inference, megakernel, kernel fusion, GPU performance modeling}}

\begin{document}
\maketitle
\begin{abstract}

Token-and-duration transducer (TDT) models enable accurate speech transcription, but translating their model-level efficiency into high-performance GPU serving requires execution strategies tailored to the distinct demands of encoding and decoding. We present an optimized inference engine for Parakeet TDT 0.6B on an NVIDIA H100 GPU. Our stage-specific engine combines row-packed FP8 encoding, shape-specialized fusion and persistent megakernel decoding without per-step launches or host synchronization. It overlaps front-end processing with encoding and applies accuracy-gated frame merging with sequence-length restoration and selective pruning without end-to-end fine-tuning. Evaluated on Parakeet TDT 0.6B at 1,024 utterances per batch, the engine transcribes 50,058 seconds of audio per second, 14 times the throughput of the evaluated TensorRT baseline, with a lower mean word error rate (WER) and an 8.9-fold reduction in GPU energy per audio second.
\end{abstract}


\section{Introduction}

Token-and-Duration Transducers (TDT) \citep{ref1} on a FastConformer encoder \citep{ref2} are popular transcription model architectures. The cost of serving is critical for production use.

We optimize Parakeet TDT 0.6B for GPU serving. Our approach combines row-packed FP8 encoding, fused kernels, persistent decoding and accuracy-gated compute reduction.

\begin{figure}[htb]
  \centering
  \includegraphics[width=2.520in]{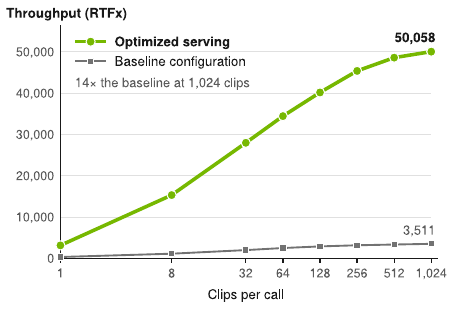}
  \caption{Throughput on an H100 SXM GPU.}
  \label{fig:throughput}
\end{figure}

On one H100, our optimized engine transcribes 50,058 seconds of audio per second (RTFx) at 1,024 utterances per call compared with 3,511 RTFx for the baseline configuration.

Our contributions include:

\begin{itemize}
  \item A row-packed FP8 encoder with two specialized GEMM designs and a megakernel for greedy TDT decoding.
  \item A training-free method for frame merging and restoration.
\end{itemize}

\section{Industrial context}

\subsection{TDT-based ASR models \& Parakeet TDT 0.6B}

A transducer predicts tokens from acoustic features and token history; TDT also predicts frame advances. Greedy TDT decoding skips frames, but differing advances across utterances make batched decoding a data-dependent loop.

\begin{figure}[htb]
  \centering
  \includegraphics[width=2.436in]{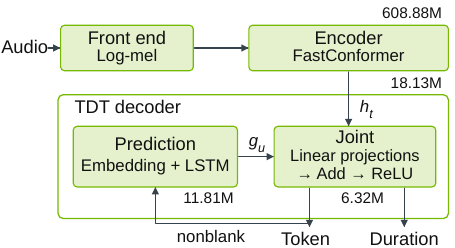}
  \caption{Parakeet TDT model architecture with parameter counts.}
  \label{fig:architecture}
\end{figure}

We target Parakeet TDT 0.6B (Figure~\ref{fig:architecture}), whose English v2 checkpoint ranked \#1 on the Hugging Face Open ASR Leaderboard in June 2025. It serves two weight sets: v2 (for English with 1,024 tokens) and v3 (for 25 European languages, automatic language detection and 8,192 tokens) \citep{ref3}.

\subsection{Target workload and serving requirements}

We target offline transcription with TDT-based transcription models. For this, the focus is to maximize peak RTFx. The peak RTFx is observed at higher batch sizes such as 1,024.

In a cascaded speech-to-speech (S2S) pipeline using TDT-based transcription models, the focus is latency at batch sizes of up to 128 requests.

\subsection{Deployment}

The baseline is deployed as the NVIDIA Parakeet TDT 0.6B NIM container \citep{ref4,ref5} (\nolinkurl{nvcr.io/nim/nvidia/parakeet-0.6b-tdt:1.3.0}) pinned to digest \texttt{sha256:\allowbreak{}0a5b5246\allowbreak{}146d0cc9\allowbreak{}e6421812\allowbreak{}dd535e64\allowbreak{}081e7a62\allowbreak{}a92c0436\allowbreak{}563ad7bf\allowbreak{}e227df06}. Both systems use the same serving stack and request concurrency. NIM divides audio into 16-second chunks with a maximum encoder execution batch size of 64; the optimized engine processes whole utterances and sizes encoder passes by valid-frame count rather than a fixed 64-utterance limit.

\section{Optimized serving}

Guided by the boundary-selection principles in \citet{ref11}, we retain separate execution regions for the log-mel front end, fused FastConformer encoder and persistent TDT decoder.

\subsection{Engine and pipeline}

\begin{figure}[htb]
  \centering
  \includegraphics[width=2.295in]{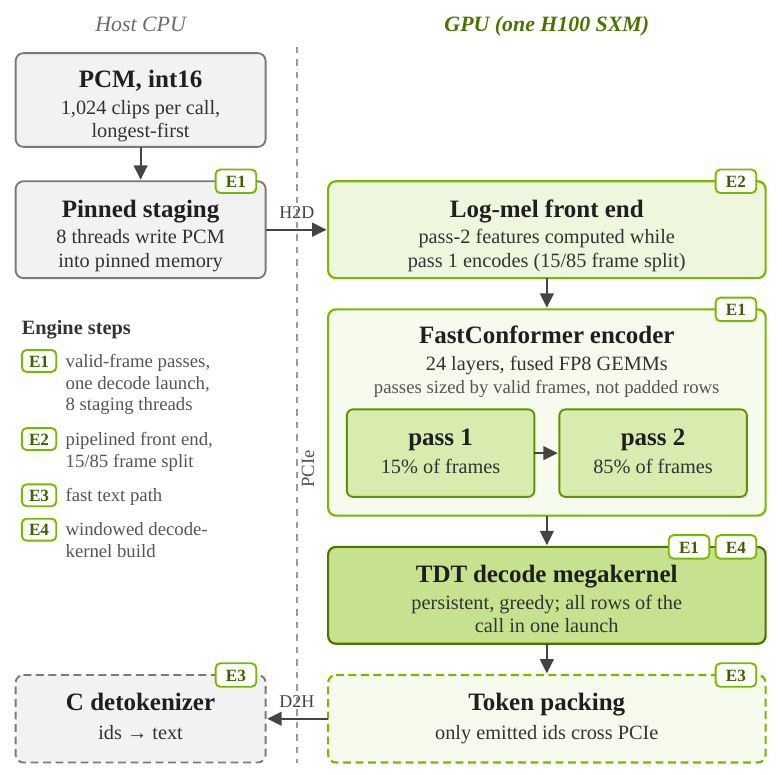}
  \caption{Host and GPU stages of one transcription call (E1--E4).}
  \label{fig:pipeline}
\end{figure}

The engine introduces three GPU libraries behind C ABIs for the three execution regions (Figure~\ref{fig:pipeline}).

\textbf{Hiding the front end (E2)}: E2 hides front-end work (INT16 PCM transfer and log-mel features) behind encoder execution on a separate CUDA stream. Calls use a 15\% / 85\% frame split. While pass 1 encodes, host threads stage pass 2 and its front end executes, leaving only pass 1's front end exposed. Exposed front-end time falls from 5.7\% to 2.1\%, increasing throughput by 6.9\%; a single large pass provides only 1.2\% because it offers insufficient overlap.

\textbf{Planning passes by frames (E1)}: The split relies on the planner. It sizes encoder passes by valid frames rather than by rows times a padded length. A 1,024-clip call then needs 1.4 passes instead of 15.9 padded 64-clip passes. The persistent decode megakernel runs over all 1,024 rows in a single launch after the last pass. This beats 256-row decode groups overlapped with later passes which take SMs from the encoder. Eight staging threads write PCM into pinned buffers, which cut host time from 9.7\% to 5.4\% of the call.

\textbf{E3} introduces a smaller copy by packing each row's actual tokens. 0.4 MB of IDs crosses PCIe per call instead of the full 1,024\ensuremath{\times}4,810 token array (19.7 MB). E3 introduces a faster detokenizer as well by using lookup tables built from SentencePiece's decoding of every token ID. \textbf{E4} selects the windowed build of the decode megakernel (+0.5\%).

\begin{figure*}[t]
  \centering
  \includegraphics[width=5.568in]{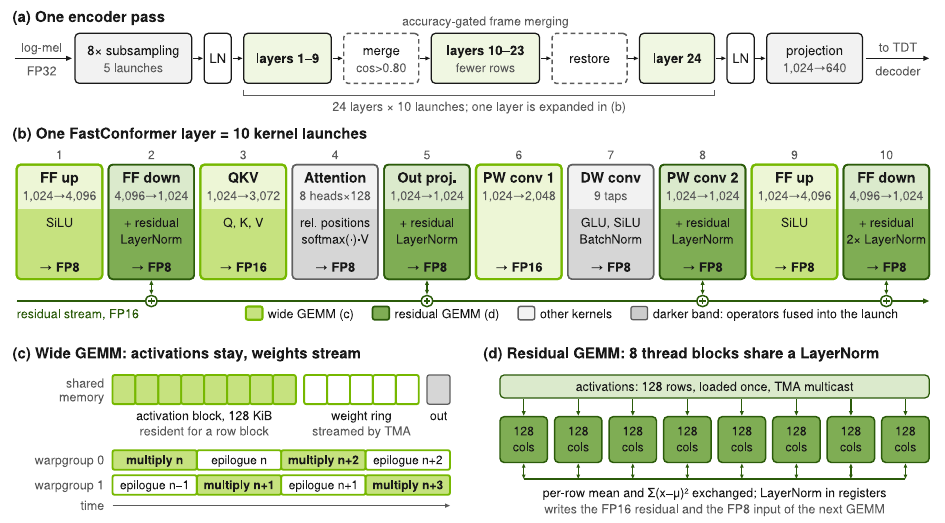}
  \caption{FastConformer encoder and FP8 GEMM designs.}
  \label{fig:encoder}
\end{figure*}

\subsection{Encoder}

The FastConformer encoder takes 80\% of wall time in the baseline configuration. Its design follows four decisions:

(1) \textbf{No padding}: Utterances differ in length; to avoid computing padding, a pass is one row-packed matrix, one row per 80 ms encoder frame.

(2) \textbf{FP8 matrix multiplies in two shape-specific kernels}: 8-bit floating-point (FP8) operands halve the operand bytes and double the peak tensor-core rate against FP16.

(3) \textbf{No extra memory pass}: LayerNorm, residual addition, activation functions and quantization are fused into the GEMM, attention and convolution kernels so the activations make no further round trip between HBM and SMs.

(4) \textbf{Less compute}: GEMM and convolution time is proportional to the row count; rows, feed-forward units and attention heads are removed under a maximum increase of 0.20 percentage points in unweighted mean development-set WER relative to the same optimized engine with merging and pruning disabled. Configurations are frozen before test evaluation.

\textbf{Precision }(Figure~\ref{fig:encoder}a, b). Five launches subsample the packed log-mel features 8\ensuremath{\times} in time to rows of width 1,024. Padding remains where the row total is rounded up to a multiple of 128 (0.13\% of rows) and where attention rounds each utterance up to blocks of 64 rows (28\% of its query rows in the encoder-only benchmark). Layer GEMMs take FP8 (E4M3) weights and activations with FP32 accumulation, one weight scale per output channel and one static activation scale per layer and GEMM; every kernel that feeds a GEMM writes FP8 already divided by that GEMM's activation scale. The residual stream is FP16 instead of FP32, which halves the bytes that each residual addition and LayerNorm reads and writes.

\textbf{Ten launches per layer (Figure~\ref{fig:encoder}(b))}: eight GEMMs, attention, and depthwise convolution. GEMMs post-process 128\ensuremath{\times}128 output tiles in epilogues. Residual-writing GEMMs (launches 2, 5, 8, 10) fuse residual addition and the following LayerNorm, producing FP16 residuals and FP8 next-GEMM inputs from the same registers. Launch 10 also computes the next layer's first LayerNorm, except in layer 24 or immediately before merging or restoration.

\textbf{Wide GEMMs}: (Figure~\ref{fig:encoder}c) FF up, QKV projection and PW conv 1 have 1,024 inputs and 2,048 to 4,096 outputs. A standard tiled kernel reloads a slice of the activations (operand A) for every output tile, although one 128-row block of A (128 KiB) feeds up to 32 tiles and fits in the shared memory of an SM. Our A-stationary kernel runs one thread block per SM over a contiguous range of output tiles and keeps the A block resident for all tiles of a row block in its range; only the weights move, through a ring of 16 KiB slots filled by TMA. Two warpgroups alternate so the epilogue of one tile overlaps the multiply of the next.

\textbf{Residual GEMMs}: (Figure~\ref{fig:encoder}d) LayerNorm requires whole-row statistics, so residual GEMMs use eight-block clusters covering the same 128 rows, each block computing 128 of 1,024 columns. A is loaded once and multicast by TMA. In the epilogue, blocks retain results in registers, exchange per-row partial means and sums of squared deviations with all eight blocks through asynchronous shared-memory stores, combine statistics, and normalize. Each cluster must fit within one hardware SM group; the CUDA runtime reports at most 15 concurrent clusters (120 of 132 SMs).

\textbf{Attention and depthwise convolution} (launches 4 and 7 in Figure~\ref{fig:encoder}b). Relative-position attention is one persistent megakernel whose thread blocks pull work items (up to 128 consecutive queries of one utterance, one head) from a shared counter. An utterance attends over its own rows only, so a pass need not be sorted by length; the padding rows of its last 64-row block are computed and discarded (queries) or masked (keys). For queries $q_i$, keys $k_j$, positional vectors $p_{i-j}$ and learned biases $u$ and $v$, we write the score as \(\operatorname{score}(i,j) = (q_i+u)\cdot k_j \allowbreak{}+ (q_i+u)\cdot p_{i-j} \allowbreak{}+ (v-u)\cdot p_{i-j}\).

The queries are loaded once for both products; the query-independent final term is precomputed per head and distance when loading the model.

DW conv is a weighted sum over nine neighboring rows (taps); taps that fall on another utterance are skipped, which is equivalent to zero-padding each utterance.

\textbf{Accuracy-gated reductions}. Feed-forward pruning removes hidden units that the preceding LayerNorm pushes so far below zero that SiLU maps them to almost zero; the rule uses the weights only and removes 10\% of the units. Head pruning removes 24 of the 192 attention heads, chosen greedily with a ridge re-fit of the output projection. We use 174 FLEURS+LibriSpeech utterances, disjoint from testing, for head selection, FP8 scale calibration and the 0.20-point WER gate; attention-output projection refitting uses ridge regularization with \ensuremath{\lambda}=2.

\begin{figure}[!b]
  \centering
  \includegraphics[width=2.308in]{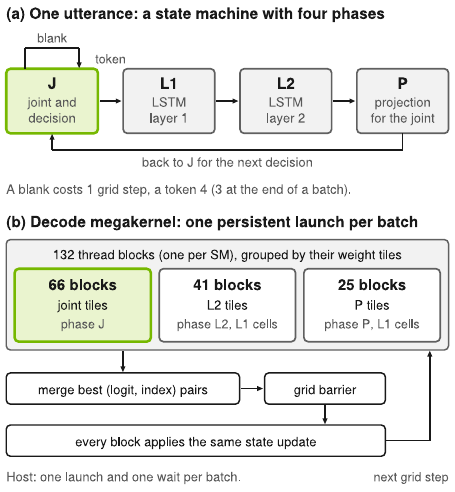}
  \caption{The decode megakernel. (a) Each utterance is a state machine. (b) One persistent launch decodes a batch. Green marks the joint, the slowest stage in grid steps.}
  \label{fig:decoder}
\end{figure}

\textbf{Frame merging} (Figure~\ref{fig:encoder}a, dashed; cf.\ \citealp{ref7}). At the merge point, adjacent rows of an utterance whose cosine similarity on the residual stream exceeds a threshold \ensuremath{\tau} are merged into their mean following Algorithm~\ref{alg:merging}. Between layers 23 and 24, each merged row is copied back to all original positions assigned to it by the mapping g, because the engine sizes the decoder input from the audio length. Passes of 42,000 rows or more merge between layers 9 and 10 with \ensuremath{\tau} = 0.80 and K=INF; shorter passes, which include the first pass of every batch, merge directly after subsampling with \ensuremath{\tau} = 0.95 and K=INF, because the host waits for the merged row count and, with a merge after layer 9, the GPU idles between passes.

\par\noindent\begin{minipage}{\columnwidth}
\begingroup
\fontsize{10}{12}\selectfont
\hrule height .7pt\kern 2.5pt
\refstepcounter{algorithm}\label{alg:merging}
\noindent\textbf{Algorithm \thealgorithm: Row Merging with Lazy Updates}\par\nopagebreak
\kern 2pt\hrule height .4pt\kern 2.5pt\nopagebreak
\noindent\hangindent=0pt\hspace*{0pt}\textbf{Input}:  rows x[1..n], threshold \ensuremath{\tau}, merge limit K (number of merges)\par
\noindent\hangindent=0pt\hspace*{0pt}\textbf{Output}: merged rows, map g from original row to merged row\par
\noindent State: \textbf{H}: hash map; H[p] = similarity of row p with its right neighbor.\quad\textbf{M}: max-heap of entries (S, a, b): similarity S of neighboring rows a, b.\quad\textbf{T}: set of merged rows whose similarities are out of date.\par
\noindent Definitions: \textbf{neighbors}: nearest active rows on each side; left(a) = left neighbor of a.\enspace An \textbf{entry (S, a, b)} is stale if a or b is inactive or H[a] \ensuremath{\ne} S.\par
\noindent Steps:\par\nopagebreak
\algline{1}{0}{mark every row active; T \ensuremath{\leftarrow} \ensuremath{\emptyset}; k \ensuremath{\leftarrow} 0}
\algline{2}{0}{for i \ensuremath{\leftarrow} 1..n\ensuremath{-}1: SCORE(i, i+1)}
\algline{3}{0}{while k \ensuremath{<} K and (M \ensuremath{\ne} \ensuremath{\emptyset} or T \ensuremath{\ne} \ensuremath{\emptyset}):}
\algline{4}{1}{e = POP-MAX(M), or none if M = \ensuremath{\emptyset}}
\algline{5}{1}{if e = none or e is stale:}
\algline{6}{2}{for each pair of neighbors (p, q) with p \ensuremath{\in} T or q \ensuremath{\in} T: SCORE(p, q)}
\algline{7}{2}{T \ensuremath{\leftarrow} \ensuremath{\emptyset}}
\algline{8}{1}{else: (S, a, b) = e, merge b into a}
\algline{9}{2}{x[a] \ensuremath{\leftarrow} (x[a] + x[b]) / 2; mark b inactive}
\algline{10}{2}{mark H[a] and H[left(a)] stale; T \ensuremath{\leftarrow} T \ensuremath{\cup} \{a\}; k \ensuremath{\leftarrow} k + 1}
\algline{11}{0}{return the active rows in order, and g[i] = number of active rows among 1..i}
\noindent \textbf{SCORE}(p, q):  H[p] \ensuremath{\leftarrow} cos(x[p], x[q]);  if H[p] \ensuremath{>} \ensuremath{\tau}: PUSH(M, (H[p], p, q))\par
\nobreak\kern 2.5pt\hrule height .7pt
\endgroup
\end{minipage}\par

\subsection{Megakernel for Greedy TDT decoding}

The decoder is a Token-and-Duration Transducer (TDT): a prediction network (token embedding and a two-layer LSTM) and a joint that adds a projection of the encoder row to a projection of the LSTM output, applies ReLU and maps the sum to logits for V = 8,192 tokens, the blank and five durations. Greedy decoding takes the best token and the best duration, moves forward by that many rows and feeds a non-blank token back into the LSTM.

One persistent megakernel decodes all 1,024 utterances across 132 SMs using internal tile scheduling (Figure~\ref{fig:decoder}), requiring one host launch and wait per batch.

Each utterance follows four grid-synchronized phases (Figure~\ref{fig:decoder}): joint scoring/decision (J), LSTM updates (L1, L2), and prediction projection (P). Blocks process their phase's active utterances, then synchronize through a global-counter barrier. Blanks remain in J for one step; tokens require four; completed utterances drop out.

Blocks retain part of their fixed weights in shared memory while active utterances change \citep{ref6}. On v3, 66, 41, and 25 blocks handle joint-output, L2, and P weights, respectively, allowing parallel stages. Folding token embeddings and L1's input matrix into a last-token lookup table, computing next-token recurrent products as soon as states exist, and caching projected LSTM outputs limit each phase to one matrix product and eliminate prediction work for blanks. Joint blocks reduce token and duration logits to (logit, index) maxima, combine them by global atomic maxima, and apply identical updates to local state copies after the barrier; no host intervention or interblock state exchange is needed.

When no confidence scores are requested, the joint skips its softmax sums and multiplies four weight tiles at a time instead of two, and the L1 cells run only on the blocks of L2 and P. On the recorded decoder inputs of 74 batches (72,341 utterances), it needs 21.4\% less decode time with tokens, frame indices, durations and token logits bitwise equal. End to end, it shortens decoding from 38.5 to 30.0 ms per batch and raises throughput by 4.3\% with identical scores.

\section{Evaluation}

\subsection{Setup and baseline}

\textbf{Hardware and model}: All measurements use one NVIDIA H100 SXM (80 GB, 700\,W, CUDA 13.2). We evaluate both weight sets of Parakeet TDT 0.6B: v3 multilingual and v2 English.

\textbf{Test data}: We use three collections. v3 is evaluated on FLEURS \citep{ref9} in 25 languages with LibriSpeech \citep{ref10} test-clean and test-other (19,185 clips, 53 hours) and on seven English test sets of the Open ASR Leaderboard \citep{ref8} (AMI, Earnings22, GigaSpeech, LibriSpeech test-clean and test-other, SPGISpeech, VoxPopuli; 72,341 clips, 158 hours). v2 is evaluated on LibriSpeech test-clean and test-other with the English set Bias (10,379 clips, 26 hours).

\textbf{Protocol}. Within each test set, the clips are sorted by descending duration and transcribed in calls of B clips. Throughput is RTFx, the inverse real-time factor: the audio duration of a collection divided by the summed time of its calls.

Timing spans request submission to final response for the baseline and optimized engine. We use five warm-up calls and three measured passes. We report mean per-pass RTFx and p95 over measured calls.

GPU energy uses NVML counter differences over measured passes, without idle-power subtraction. WER comparisons use paired-bootstrap 95\% confidence intervals with 1,000 resamples.

\subsection{Optimized serving}

Table~\ref{tab:throughput} reports 14x, 17x, and 15x throughput gains for its three collections. At batch size 128, FLEURS+LibriSpeech p95 falls from 664.3 to 55.0 ms, enabling cascaded disaggregated S2S with existing LLM, TTS, and turn-taking models at batch size 128 by limiting end-to-end latency to 1 second.

Across 1 to 1,024 clips per call (Figure~\ref{fig:throughput}), gains are 13 to 17x from eight clips upward and 9 to 10x for one clip, whose p95 is 3.6 to 3.9 ms versus 36 to 52 ms. At 1,024 clips, energy falls from 0.1204\,J to 0.0135\,J per audio second.

\par\addvspace{4pt}
\noindent\begin{minipage}{\columnwidth}
\captionsetup{type=table,skip=4pt}
\centering
\normalfont\fontsize{10}{11.5}\selectfont
\setlength{\tabcolsep}{1.15pt}
\setlength{\arrayrulewidth}{.5pt}
\begin{tabularx}{\columnwidth}{|>{\raggedright\arraybackslash}X|c|c|c|c|}
\hline
\rowcolor[HTML]{DAF2D0}
& \multicolumn{2}{c|}{\textbf{RTFx}} & \multicolumn{2}{c|}{\textbf{WER (\%)}} \\
\hhline{|~|--|--|}
\rowcolor[HTML]{DAF2D0}
\multirow{-2}{=}{\raggedright\textbf{Model, data}} &
Baseline & Optimized & Baseline & Optimized \\
\hline
v3, \mbox{FLEURS +} \mbox{LibriSpeech} & 3,511 & 50,058 & 7.86 & 6.56 \\
\hline
v3, Open ASR & 2,878 & 48,736 & 6.06 & 5.12 \\
\hline
\mbox{v2, Bias +} \mbox{LibriSpeech} & 3,258 & 48,465 & 3.35 & 1.99 \\
\hline
\end{tabularx}
\caption{Throughput and mean WER at 1,024 clips on one H100 SXM GPU.}
\label{tab:throughput}
\end{minipage}\par

\par\addvspace{4pt}
\noindent\begin{minipage}{\columnwidth}
\captionsetup{type=table,skip=4pt}
\centering
\normalfont\fontsize{10}{11.5}\selectfont
\setlength{\tabcolsep}{1.15pt}
\setlength{\arrayrulewidth}{.5pt}
\begin{tabularx}{\columnwidth}{|>{\raggedright\arraybackslash}X|c|c|c|c|}
\hline
\rowcolor[HTML]{DAF2D0}
& \multicolumn{2}{c|}{\textbf{\shortstack{Completion\\time p95}}} & \multicolumn{2}{c|}{\textbf{\shortstack{Energy per\\audio second}}} \\
\hhline{|~|--|--|}
\rowcolor[HTML]{DAF2D0}
\multirow{-2}{=}{\raggedright\textbf{Model, data}} &
Baseline & Optimized & Baseline & Optimized \\
\hline
v3, \mbox{FLEURS +} \mbox{LibriSpeech} & 664.3 ms & 55.0 ms & 0.1103 J & 0.0152 J \\
\hline
v3, Open ASR & 458.9 ms & 43.3 ms & 0.1292 J & 0.0158 J \\
\hline
\mbox{v2, Bias +} \mbox{LibriSpeech} & 628.9 ms & 58.1 ms & 0.1136 J & 0.0157 J \\
\hline
\end{tabularx}
\caption{95th-percentile completion time and energy consumed per audio second at 128 clips on one H100 SXM GPU.}
\label{tab:latency}
\end{minipage}\par

The optimized serving has the lower mean WER on all three collections, by 0.9 to 1.4 points. On FLEURS, it is significantly better in 19 of 25 languages and worse in none. Avoiding chunking explains the WER improvement over the baseline; merging and pruning increase mean WER by 0.12, 0.16 and 0.05 percentage points relative to our engine without these reductions.

The optimized serving draws up to 680\,W against 425\,W for the baseline but reduces GPU energy per audio second by a factor of up to 9. It holds up to 7\,GiB more GPU memory.

\section{Optimizations ruled out and learnings}

\textbf{Encoder as a monolithic megakernel}. A single persistent kernel for all 24 layers was 0.69 times as fast (185.4 against 128.7 ms per pass). The loss was not in scheduling: with a barrier after every operator, the kernel was still only 0.64 times as fast. Its matrix-multiply loop waited for operands during up to 46\% of its time, and its fastest task order was operator by operator, the order of the chain. Unlike the decoder, an encoder pass is bound by compute and operand traffic which a monolithic megakernel does not reduce. This result is consistent with observations in Nemotron 3.5 Transcribe serving \citep{ref12}, where fused kernels outperform persistent megakernels for larger encoder passes. This makes the encoder megakernel-averse \citep{ref11}.

\textbf{Gains that did not survive}: A GEMM epilogue with a deeper pipeline kept about one seventh of its isolated gain in a whole pass. Frame merging in the middle of the stack saved 3.5 ms per pass on the encoder-only benchmark and lost 2.4 to 4.2\% end-to-end as the host waited for the merged row count and the GPU idled between passes. Overlapping decoding with the next encoder pass lost 0.8 to 2.9\%: the decode kernel occupies all SMs, so overlap on the GPU is zero-sum.

\section{Conclusion}

Our results show that fusion must match the workload: persistent execution benefits sequential decoding while specialized fused kernels better serve the compute-intensive encoder. This establishes a framework for optimization for megakernel-averse architectures.

{\fontsize{10}{12}\selectfont
\bibliography{references}

@inproceedings{ref1,
  author    = {Hainan Xu and Fei Jia and Somshubra Majumdar and He Huang and Shinji Watanabe and Boris Ginsburg},
  title     = {Efficient Sequence Transduction by Jointly Predicting Tokens and Durations},
  booktitle = {Proceedings of the 40th International Conference on Machine Learning},
  year      = {2023},
  series    = {Proceedings of Machine Learning Research},
  volume    = {202},
  pages     = {38462--38484},
  publisher = {PMLR},
  url       = {https://proceedings.mlr.press/v202/xu23g.html}
}

@inproceedings{ref2,
  author    = {Dima Rekesh and Nithin Rao Koluguri and Samuel Kriman and Somshubra Majumdar and Vahid Noroozi and He Huang and Oleksii Hrinchuk and Krishna Puvvada and Ankur Kumar and Jagadeesh Balam and Boris Ginsburg},
  title     = {Fast {Conformer} with Linearly Scalable Attention for Efficient Speech Recognition},
  booktitle = {IEEE Automatic Speech Recognition and Understanding Workshop (ASRU)},
  year      = {2023},
  publisher = {IEEE},
  doi       = {10.1109/ASRU57964.2023.10389701},
  url       = {https://ieeexplore.ieee.org/document/10389701/}
}

@misc{ref3,
  author        = {Monica Sekoyan and Nithin Rao Koluguri and Nune Tadevosyan and Piotr Zelasko and Travis Bartley and Nikolay Karpov and Jagadeesh Balam and Boris Ginsburg},
  title         = {{Canary-1B-v2} \& {Parakeet-TDT-0.6B-v3}: Efficient and High-Performance Models for Multilingual {ASR} and {AST}},
  year          = {2025},
  eprint        = {2509.14128},
  archiveprefix = {arXiv},
  primaryclass  = {cs.CL},
  url           = {https://arxiv.org/abs/2509.14128}
}

@misc{ref4,
  author       = {{NVIDIA}},
  title        = {{Parakeet TDT 0.6b En-US}, version 1.3.0},
  year         = {n.d.},
  howpublished = {{NVIDIA NGC}, {NIM} container},
  note         = {Accessed November 24, 2026},
  url          = {https://catalog.ngc.nvidia.com/orgs/nim/nvidia/containers/parakeet-0.6b-tdt/1.3.0}
}

@misc{ref5,
  author       = {{NVIDIA}},
  title        = {{Parakeet TDT}: Deployment Guide},
  year         = {2026},
  howpublished = {{NVIDIA Speech NIM Microservices} documentation, release 26.02.0},
  note         = {Accessed November 24, 2026},
  url          = {https://docs.nvidia.com/nim/speech/26.02.0/asr/deploy-asr-models/parakeet-tdt.html}
}

@inproceedings{ref6,
  author    = {Greg Diamos and Shubho Sengupta and Bryan Catanzaro and Mike Chrzanowski and Adam Coates and Erich Elsen and Jesse Engel and Awni Hannun and Sanjeev Satheesh},
  title     = {Persistent {RNNs}: Stashing Recurrent Weights On-Chip},
  booktitle = {Proceedings of the 33rd International Conference on Machine Learning},
  year      = {2016},
  series    = {Proceedings of Machine Learning Research},
  volume    = {48},
  pages     = {2024--2033},
  publisher = {PMLR},
  url       = {https://proceedings.mlr.press/v48/diamos16.html}
}

@inproceedings{ref7,
  author    = {Daniel Bolya and Cheng-Yang Fu and Xiaoliang Dai and Peizhao Zhang and Christoph Feichtenhofer and Judy Hoffman},
  title     = {Token Merging: Your {ViT} But Faster},
  booktitle = {International Conference on Learning Representations (ICLR)},
  year      = {2023},
  url       = {https://openreview.net/forum?id=JroZRaRw7Eu}
}

@misc{ref8,
  author        = {Vaibhav Srivastav and Steven Zheng and Eric Bezzam and {Le Bihan}, Eustache and Nithin Rao Koluguri and Piotr {\.Z}elasko and Somshubra Majumdar and Adel Moumen and Sanchit Gandhi},
  title         = {Open {ASR} Leaderboard: Towards Reproducible and Transparent Multilingual and Long-Form Speech Recognition Evaluation},
  year          = {2025},
  eprint        = {2510.06961},
  archiveprefix = {arXiv},
  primaryclass  = {cs.CL},
  url           = {https://arxiv.org/abs/2510.06961}
}

@inproceedings{ref9,
  author    = {Alexis Conneau and Min Ma and Simran Khanuja and Yu Zhang and Vera Axelrod and Siddharth Dalmia and Jason Riesa and Clara Rivera and Ankur Bapna},
  title     = {{FLEURS}: Few-shot Learning Evaluation of Universal Representations of Speech},
  booktitle = {2022 IEEE Spoken Language Technology Workshop (SLT)},
  year      = {2023},
  pages     = {798--805},
  publisher = {IEEE},
  doi       = {10.1109/SLT54892.2023.10023141},
  url       = {https://ieeexplore.ieee.org/document/10023141/}
}

@inproceedings{ref10,
  author    = {Vassil Panayotov and Guoguo Chen and Daniel Povey and Sanjeev Khudanpur},
  title     = {{LibriSpeech}: An {ASR} Corpus Based on Public Domain Audio Books},
  booktitle = {IEEE International Conference on Acoustics, Speech and Signal Processing (ICASSP)},
  year      = {2015},
  pages     = {5206--5210},
  publisher = {IEEE},
  doi       = {10.1109/ICASSP.2015.7178964},
  url       = {https://ieeexplore.ieee.org/document/7178964}
}

@misc{ref11,
  author        = {Aditya Chatterjee and Anand Joseph X M and Sanjay Singh Chauhan},
  title         = {One Megakernel is Not Enough: Predicting Where to Cut for Streaming Inference},
  year          = {2026},
  eprint        = {2610.35336},
  archiveprefix = {arXiv},
  primaryclass  = {cs.PF},
  url           = {https://arxiv.org/abs/2610.35336}
}

@misc{ref12,
  author        = {Aditya Chatterjee and Sanjay Singh Chauhan and Mayank Jain and Arushi Dhingra and Sahin Chowdhury and Rahul Mittal},
  title         = {High-Throughput Serving of {Nemotron} 3.5 Transcribe with Regime-Aware Megakernels and Kernel Fusion},
  year          = {2026},
  eprint        = {2611.20095},
  archiveprefix = {arXiv},
  primaryclass  = {cs.PF},
  url           = {https://arxiv.org/abs/2611.20095}
}
}
\end{document}